\documentclass[aps,prx,twocolumn,groupeaddress,superscriptaddress]{revtex4-2}

\usepackage[dvipdfmx]{graphicx}
\usepackage{helvet}
\usepackage{amsmath,amssymb}
\usepackage{bm}
\usepackage{colortbl}
\usepackage{braket}
\begin{document}


\title{Microscopic Evidence for Preformed Cooper Pairs in Pressure-Tuned Organic Superconductors near Mott Transition}

\author{Tetsuya Furukawa}
\email{tetsuya.furukawa.c1@tohoku.ac.jp}
\affiliation{Department of Applied Physics, University of Tokyo, Tokyo 113-8656, Japan}
\affiliation{Institute for Materials Research, Tohoku University, Sendai 980-8577, Japan}

\author{Kazuya Miyagawa}
\affiliation{Department of Applied Physics, University of Tokyo, Tokyo 113-8656, Japan}

\author{Mitusnori Matsumoto}
\affiliation{Department of Applied Physics, University of Tokyo, Tokyo 113-8656, Japan}

\author{Takahiko Sasaki}
\affiliation{Institute for Materials Research, Tohoku University, Sendai 980-8577, Japan}

\author{Kazushi Kanoda}
\email{kanoda@ap.t.u-tokyo.ac.jp}
\affiliation{Department of Applied Physics, University of Tokyo, Tokyo 113-8656, Japan}

\date{\today}

\begin{abstract}
{
A weird electronic state accompanied with an anomalous superconducting precursor and/or exotic orders, called the pseudogap state, arises prior to a superconducting condensate in underdoped cuprates that are situated near Mott transition. 
Another way to make the system approach the Mott transition is the variation of bandwidth or correlation strength, which gives a new dimension to exploring this exotic state.
Here we report nuclear magnetic resonance (NMR) studies on layered organic superconductors with half-filled bands whose widths are pressure-tuned near the Mott transition.
The system situated on the verge of the Mott transition shows a pseudogap-like anomalous suppression of spin excitations on cooling from well above the superconducting critical temperature $T_{\mathrm{c}}$.
The pressure variation of the NMR relaxation rate shows that the pseudogap-like behavior is rapidly suppressed by applying pressure.
The NMR experiments under various magnetic fields varied up to 18 T proves the absence of symmetry breaking orders that compete with superconductivity, such as charge orders, in the metallic phase.
Remarkably, the pseudogap-like behavior above $T_{\mathrm{c}}$ and the superconducting condensate fade out in parallel under ascending magnetic fields with similar field-orientation dependence, indicating a superconducting precursor is the predominant origin of the pseudogap.
Our further investigation of different materials, which take different “distances” from the Mott transition by chemical pressure, confirms that the superconducting precursor is not the conventional amplitude fluctuations arising from low dimensionality but unconventional preformation of Cooper pairs enhanced near the Mott transition.
These findings conclude that preformed Cooper pairs persist up to twice as high as $T_{\mathrm{c}}$ on the verge of the bandwidth-controlled Mott transition.
}
\end{abstract}


\maketitle
\section{Introduction}

\begin{figure*}[!th]
\begin{center}
\includegraphics[width=16 cm,clip]{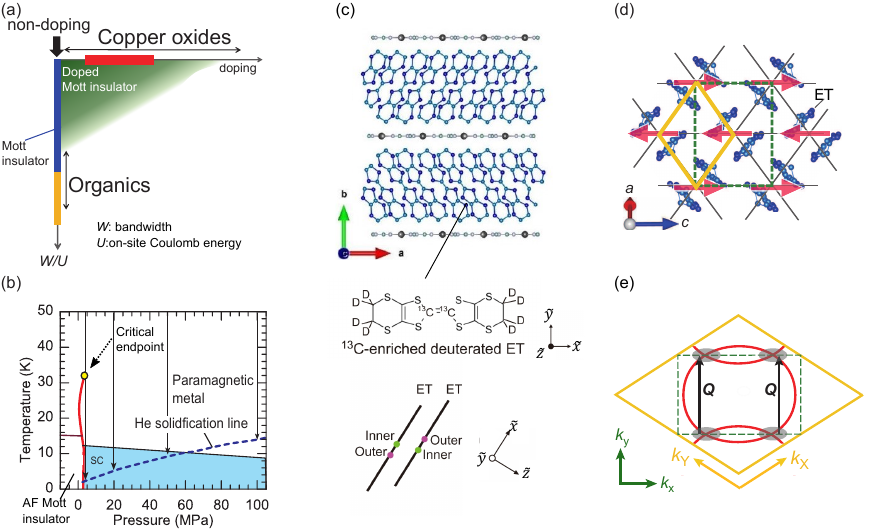}
\caption
{
(Color online) Basic properties of the organic superconductor $\kappa$-(ET)$_{2}$$X$. (a) Comparison of $\kappa$-(ET)$_{2}$$X$ and copper oxides on the parameter plane spanned by doping level and the interaction strength. (b) Pressure-temperature phase diagram of $\kappa$-dBr. The red line represents the first-order BCMT line. The solid arrows indicate the temperature scans covered by NMR measurements under pressure. The pressure medium, helium, solidifies at the broken line, the region below which was inaccessible in the present experiments.
(c) Layered structure of $\kappa$-dBr and schematic of ET molecules enriched with $^{13}$C isotopes. The $\tilde{x}$, $\tilde{y}$, and $\tilde{z}$ axes here are principal axes of an ET molecule.
(d) Magnetic structure of AF long-range ordered $S$ = 1/2 spins with a $\bm{Q}$ = ($\pi$, $\pi$) ordering vector in the Mott insulating state of $\kappa$-dBr. Each ET dimer, which corresponds to the lattice site of an anisotropic triangular lattice, accommodates one hole. The yellow diamond and the green rectangle correspond to a primitive and an enlarged unit cell of an anisotropic triangular lattice, respectively. 
(e) Schematic Fermi surfaces of the metallic state of $\kappa$-dBr. The yellow diamond and the green rectangle are the first Brillouin zones of the unit cells with the corresponding colors in (d). The black arrows indicate a wavevector $\bm{Q}$ = ($Q$$_{X}$, $Q$$_{Y}$) = ($\pi$, $\pi$). The gray circled area corresponds to the region around which a superconducting gap of $d$$_{xy}$ (=$d$$_{X^{2}-Y^{2}}$) symmetry opens.
}
\label{Fig1} 
\end{center}
\end{figure*}

Lightly doped copper oxides show anomalous electronic and magnetic behaviors prior to the superconducting transition~\cite{Timusk1999,Hashimoto2014,Keimer2015,Uchida2021}, a phenomenon called pseudogap that has been discussed in terms of superconducting precursors~\cite{Li2010,Kondo2015a}, momentum-dependent quasiparticle coherence (Fermi-arc)~\cite{Hashimoto2014,Stanescu2006,Sakai2009}, and the emergence of various exotic orders. Although the issue is still under debate, reconstruction of Fermi surfaces~\cite{Doiron-Leyraud2007,Wu2018,Braganca2018,Fang2022} and the emergence of complex orders such as charge density wave~\cite{Ghiringhelli2012, Chang2012a, Wu2011,Uchida2021}, a time-reversal-symmetry-broken state~\cite{Xia2008}, a nematic state~\cite{Lawler2010,Sato2017a,Nie2014}, a topological order~\cite{Wu2018,Scheurer2018}, and pair density wave~\cite{Lee2014,Agterberg2019} in lightly doped systems are among the recent advances related to this issue. The pseudogap has been addressed as an issue of doped Mott insulators, where the density of doped carriers is a controlling parameter. More broadly, however, not only the density of doped carriers but also the strength of interactions among electrons is another key parameter that dominates the behavior of correlated electrons; the behavior should be explored in the parameter plane spanned by doping level and the interaction strength (Fig.~\ref{Fig1}(a)). Thus, to look into a situation contrasting to that of the copper oxides, namely, without doping but with varying strength of the interactions is expected to open a new dimension to this issue.

A layered organic superconductor (SC), $\kappa$-(ET)$_2$$X$, where ET represents bis(ethylenedithio)-tetrathiafulvalene, with a half-filled band is in the desired situation; unconventional superconductivity ($d$-wave) ~\cite{Schrama1999,Malone2010,Izawa2001,Dion2009,Oka2015,Guterding2016,Cavanagh2019,Imajo2021} emerges from Mott insulators without doping but by varying the bandwidth~\cite{Kanoda2006,Limelette2003}, which controls the relative strength of the interactions to the kinetic energy (Fig.~\ref{Fig1}(a)). It is well recognized that the application of physical pressure and/or chemical pressure by anion $X$ substitution finely tunes the interaction strength, which is a primary factor that dominates the electronic properties near the Mott transition~\cite{Kanoda2006,Limelette2003,Kagawa2005,Furukawa2015,Furukawa2018}. In particular, $\kappa$-(ET)$_2$Cu[N(CN)$_2$]Br with all protons in ET substituted by deuterons (abbreviated to $\kappa$-dBr) is on the verge of the bandwidth-controlled Mott transition (BCMT) (Figs.~\ref{Fig1}(b) and (c)) from a Mott insulator with an antiferromagnetic (AF) order of a wavevector, $\bm{Q}$ = ($\pi$, $\pi$), (Fig.~\ref{Fig1}(d)) to a metal with a cylindrical Fermi surface~\cite{Miyagawa2002,Miyagawa2004a,Sasaki2008a} (Fig.~\ref{Fig1}(e)). As has been revealed in the study of copper oxides~\cite{Chang2012a,Wu2011,Yasuoka1989,Warren1989,Takigawa1991,Zheng1999,Bachman1999,Mitrovic1999}, iron pnictides~\cite{Grafe2008,Nakai2010a}, and iron chalcogenides~\cite{Kasahara2016}, nuclear magnetic resonance (NMR) is a useful microscopic probe to elucidate unconventional electronic states at temperatures above $T_{\rm c}$. In ordinary metals, nuclear spin-lattice relaxation rate divided by temperature, ($T_{1}T$)$^{-1}$, is independent of temperature and magnetic field because the dynamic spin susceptibility of conduction electrons does not sensitively vary with those parameters. However, the metallic phase of $\kappa$-dBr shows an anomalous decrease in ($T_{1}T$)$^{-1}$ on cooling from above $T_{\rm c}$~\cite{Miyagawa2002}, which is compared to the pseudogap in the cuprates. 
It is noted that the transport properties such as the Nernst coefficient~\cite{Nam2007} and the magnetic susceptibility~\cite{Uehara2013} suggest superconducting fluctuations enhanced near BCMT.

The present work explores the anomalous suppression of spin excitations in $\kappa$-dBr through $^{13}$C NMR under tuning interaction strength by He-gas pressure and systematically suppressing superconductivity by magnetic fields over a twenty-fold range(0.9 T to 18 T).
We found that in the non-doped case the anomalous decrease in spin excitations on cooling occurs mainly in the channel of AF fluctuations very probably with $\bm{Q}$ = ($\pi$, $\pi$), which is enhanced near the BCMT, and the anomaly fades out along with superconductivity under increasing magnetic field. 
Our further investigation of different materials, which are located in different ``distances'' from the Mot transition by chemical pressure, suggests that the phenomena cannot be explained by conventional superconducting amplitude fluctuations but is an indication of unconventional superconducting phase fluctuations due to preformation of Cooper pairs. 
We note that this observation is free from the intricate issue of competing or coexisting orders as observed in doped copper oxides, because in the measured temperates, pressures, and magnetic fields, the NMR spectra do not show any splitting, broadening or complicated structures indicative of symmetry breaking such as charge order in a metallic phase~\cite{Miyagawa2000}.
We also note that fluctuaions of nematic order, which breaks rotational symmetry and occurs in the pseudogap phase in cuprates, does not pertain to the present system, which has an orthorhombic or monoclinic crystal structure and no room for nematic order.

\begin{figure*}[!ht]
\begin{center}
\includegraphics[width=18cm,clip]{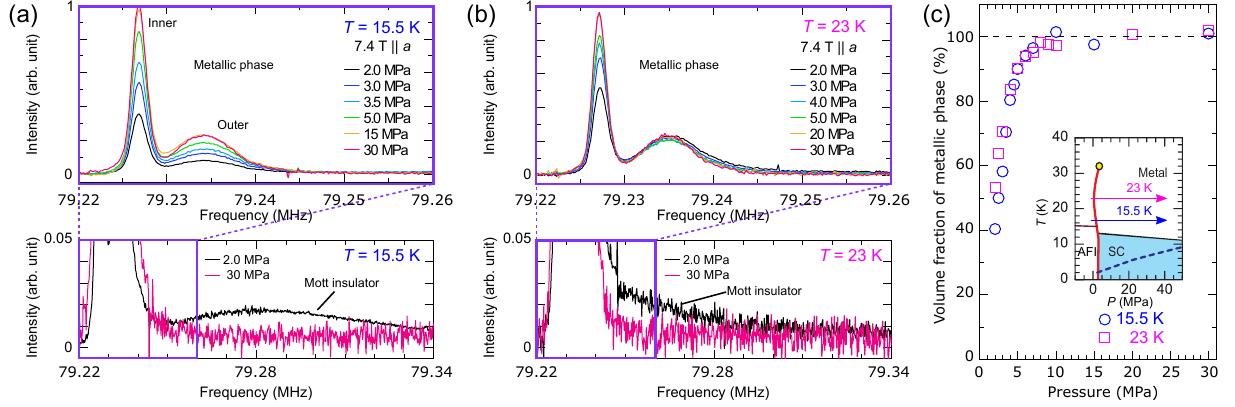}
\caption{(Color online) $^{13}$C NMR spectra of $\kappa$-dBr under pressures. (a) at 15.5 K and (b) at 23 K. A magnetic field of 7.4 T is applied parallel to the conducting layers. (c) Pressure dependence of the volume fraction of the metallic phase at 15.5 K and 23 K in $\kappa$-dBr. (inset) 
Pressure-temperature phase diagram of $\kappa$-dBr (see Fig.~\ref{Fig1}(b)). The solid arrows indicate the pressure scans covered by NMR spectral measurements.}
\label{Fig22} 
\end{center}
\end{figure*}

\section{Experimental Procedure}

The $^{13}$C NMR experiments were performed on a single crystal of $\kappa$-dBr, in which the central double-bonded carbon sites of ET molecules had been enriched with $^{13}$C isotopes with nuclear spin $I$ =1/2 and gyromagnetic ratio $\gamma$/2$\pi$ = 10.705 MHz/T (Fig.~\ref{Fig1}(c)). When an external magnetic field is applied to $\kappa$-dBr with an arbitrary orientation to the crystal axes, the $^{13}$C NMR spectra generally consist of 16 (2 $\times$ 2 $\times$ 4 as explained below) resonance lines, which have three different origins of the line splitting: (i) the shifted face-to-face dimerization of ET molecules makes the two central carbon sites in ET inequivalent (called ``inner'' and ``outer'' sites, as depicted in Fig.~\ref{Fig1}(c), giving two lines with different shifts; (ii) each line further splits into two, owing to the nuclear dipolar fields from the adjacent $^{13}$C nuclei, which is called the ``Pake doublet''; (iii) the unit cell contains four dimers (two in a layer), which are all inequivalent with respect to the magnetic field direction except for high-symmetry magnetic field orientations ( $\mu_{0}H$ $\parallel$ $a$, $\parallel$ $b$, or $\parallel$ $c$) (see Fig.~\ref{Fig1}(c)).
In the present study, magnetic fields, $\mu_{0}H$, were applied parallel to the $a$ axis (in-plane) or the $b$ axis (out-of-plane) (see Fig.~\ref{Fig1}(c)). In these situations, all of the dimers are equivalent, so the splitting of the origin (iii) does not occur; thus, the number of NMR lines is reduced to four. In particular, in the case $\mu_{0}H$ $\parallel$ $a$, the splitting from the origin (ii) is negligible because the angle between the $^{13}$C-$^{13}$C bonding direction and the magnetic field direction is close to the special angle called the magic angle, at which the dipolar splitting vanishes. Thus, only two lines are observed, and these correspond to the inner- and outer-site spectra, owing to splitting origin (i) (see Fig.~\ref{Fig22}).
As explained in detail later, the metallic phase was clearly separated from the insulating phase in spectrum even when they coexist on the verge of the Mott transition around ambient pressure. The nuclear spin-lattice relaxation rate, $T_{1}^{-1}$, was determined from the recovery curves of nuclear magnetization following the saturation comb pulses: i.e., 1-$I(t)/I(\infty)$ = $A$exp$(-t/T_{1})$, where $t$ is the recovery time, $A$ is the fitting constant, and $I(t)$ is the integrated intensity of NMR spectra at time $t$. The pressure was finely tuned near the BCMT by using He gas as the pressure medium. The details of the sample preparation, NMR experiments, and pressurizing techniques are available in Appendix~\ref{Appendix A}.

\section{Pressure study of $^{13}$C NMR}

In this section, we first show variation of the NMR spectra across the Mott transition induced by applying pressure and also demonstrate that the metallic phase can be separated from the insulating phase in spectrum on the verge of the first-order Mott transition. 
We next show how, in the metallic phase, the AF fluctuations and the pseudogap-like behavior evolve while approaching the Mott transition by pressure variation. 
Comparing pressure dependences of ($T_{1}T$)$^{-1}$ and Knight shift $K$, we conclude that the ``pseudogap'' emerge at particular $\bm{k}$-regions that affect AF spin fluctuations with $\bm{Q}$ = ($\pi$, $\pi$).

\begin{figure*}[!hbt]
\begin{center}
\includegraphics[width=17cm,clip]{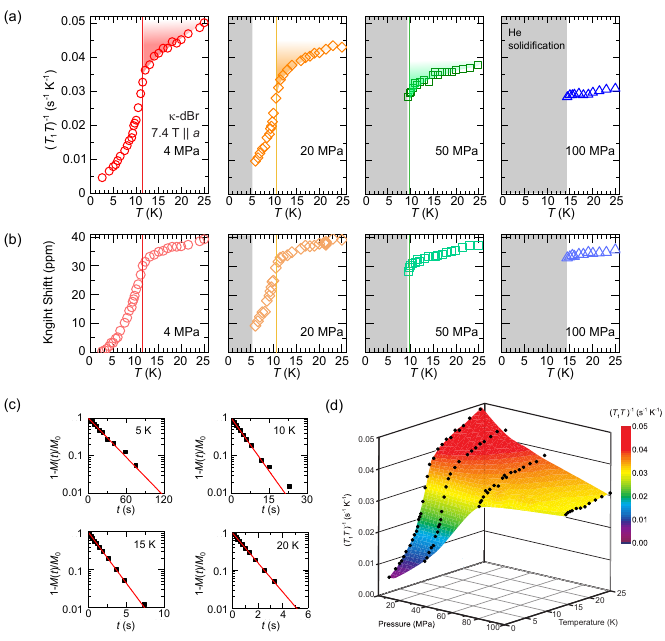}
\caption{(Color online) Pressure variation of the temperature dependences of (a) ($T_{1}T$)$^{-1}$ and (b) the Knight shift, $K$. 
The data for 20 MPa, 50 MPa and 100 MPa at lower temperatures are absent below 5.2 K, 9.1 K and 14.1 K, respectively, because the area below the He solidification line in Fig. 1(b) is experimentally inaccessible. The vertical lines indicate the $T_{\rm c}$'s at each pressure, as determined from the kinks in the Knight shift.
(c)$^{13}$C NMR relaxation curves of $\kappa$-dBr at a pressure of 4 MPa.
The recovery of nuclear magnetization $I(t)$ is plotted in the form of log 1-$I(t)/I(\infty)$ vs $t$ for several temperatures, where $t$ is the time of recovery and $I(\infty)$ is the saturated value of $I(t)$. All of the data are well fitted by straight lines: 1-$I(t)/I(\infty)$ $\propto$ exp$(-t/T_{1})$ with $T_{1}$ nuclear spin-lattice relaxation time.
(d) A contour plot of ($T_{1}T$)$^{-1}$ as a function of temperature and pressure.
}
\label{Fig23} 
\end{center}
\end{figure*}

\subsection{NMR spectra under pressure variation}
First, we show pressure dependence of NMR spectra. A magnetic field of 7.4 T was applied parallel to the $a$ axis (in-pane); hence, two lines were observed, as shown in Figs.~\ref{Fig22}(a) and (b), which indicates the pressure evolution of the NMR spectra at 15.5 K and 23 K under pressure (also see the inset of Fig.~\ref{Fig22}(c)). The two lines at 15.5 K arise from a metallic phase (upper panel in Fig.~\ref{Fig22}(a)); the line at the lower (higher) frequency is from the inner (outer) site (see also Fig.~\ref{Fig1}(c)).
At 2 MPa, a broad line appearing at a higher frequency of approximately 79.280 MHz in the lower panel of Fig.~\ref{Fig22}(a) arises from an antiferromagnetic Mott insulating phase, because the metallic and insulating phases coexist around the critical pressure of the Mott transition owing to its first-order nature; however, they are clearly separated in the NMR spectra. The pressure evolution of the volume fraction of the metallic phase, which is evaluated from the intensity of the inner-site line (approximately 79.227 MHz) of the metallic phase, is shown in Fig.~\ref{Fig22}(c). 
At 23 K, the double-peaked spectrum (the upper panel of Fig.~\ref{Fig22}(b)) is similarly from the metallic phase. As seen in the lower panel, the spectral component of the Mott-insulating phase comes close to the metallic phase. As a result, the outer line of the metallic phase possibly overlaps with the inner line of the insulating component because the inner line at 79.227 MHz rapidly develops with pressure, indicating the growth of the metallic-phase volume fraction as shown in Fig.~\ref{Fig22}(c), whereas the outer-line at 79.235 MHz appears unchanged in intensity with slight variation in shape; this is possibly because an increase in the intensity of the outer-site line in the metallic phase is compensated by the decrease in the inner-site line intensity in the Mott-insulating phase. To avoid the contamination of the Mott insulating phase albeit cautious only at low pressures, we deduced the Knight shift and the relaxation rate in the metallic and superconducting phases from the inner-site spectra in the present study.

\subsection{Pressure dependence of AF fluctuations}

Here, we show how the AF fluctuations evolve while approaching the Mott transition by applying pressure.
Figure~\ref{Fig23}(a), which is one of the main results of the present study, shows the temperature dependence of ($T_{1}T$)$^{-1}$ at the pressures indicated in Fig.~\ref{Fig1}(b). 
The lowest temperatures available, 5.2, 9.2 and 14.1 K for 20, 50 and 100 MPa, respectively, are limited by helium solidification (see Fig.~\ref {Fig1}(b)).
As mentioned above, the inner line of the metallic phase (Fig.~\ref{Fig22}(a)) is investigated to perfectly exclude the possible admixture of the insulating phase in the line investigated. 
Indeed, the relaxation curves for the inner lines for the metallic phase were single-exponential functions of time even at low pressures, confirming the single-phase (metallic-phase) nature of the lines investigated; for example, the data of 4 MPa are shown in Fig.~\ref{Fig23}(c).
Sharp decreases in ($T_{1}T$)$^{-1}$ at approximately 11 K in Fig.~\ref{Fig23}(a) are due to the superconducting transition. It is evident that ($T_{1}T$)$^{-1}$ is sensitive to pressure. The level of ($T_{1}T$)$^{-1}$ in the normal state above approximately 11 K increases while approaching the BCMT as the pressure decreases from 100 MPa to 4 MPa. Concerning the temperature dependence, ($T_{1}T$)$^{-1}$ at 4 MPa, which is very close to the BCMT, clearly decreases on cooling toward $T_{\rm c}$. When the system is driven away from the BCMT by pressure, the feature becomes less prominent and, at 100 MPa, ($T_{1}T$)$^{-1}$ is nearly temperature-independent, as expected in conventional paramagnetic metals. The pressure and temperature profiles of ($T_{1}T$)$^{-1}$ show that the anomalous suppression of spin excitations on cooling emerges near the Mott transition, where AF fluctuations develop, as visualized in Fig.~\ref{Fig23}(d). 

Compared with the remarkable temperature and pressure dependences of ($T_{1}T$)$^{-1}$, the Knight shift, $K$, which measures the uniform spin susceptibility $\chi'$($\bm{q}$ = 0), exhibits only moderate decreases with temperature and less prominent pressure dependence in the normal state, as shown in Fig.~\ref{Fig23}(b). 
The Knight shifts in Fig.~\ref{Fig23}(b) correspond to the peak frequencies of the inner-site spectra at each temperature, and the origin of the Knight shift was determined by extrapolating the peak frequencies at 4 MPa to absolute zero\footnote{The metallic phase of $\kappa$-dBr shows singlet superconductivity at low temperatures, where the Knight shift is expected to vanish at absolute zero without a residual shift owing to the negligible spin-orbit interaction in the present materials.}.
This behavior of Knight shift indicates that $\chi'$($\bm{q}$ = 0) is less coupled to the anomalous spin excitations than is the finite-$\bm{q}$ spin susceptibility probed by ($T_{1}T$)$^{-1}$. Note that the form factor of the $^{13}$C sites in the present system is nearly $\bm{q}$-independent and the so-called Korringa ratio $\mathcal{K}_{\alpha}$ $\propto (KT_{1}T)^{-1}$ estimated from the ($T_{1}T$)$^{-1}$ and $K$ values yields 8-12, which is much greater than unity, indicating that ($T_{1}T$)$^{-1}$ is dominated by finite-$\bm{q}$ spin fluctuations (Appendix~\ref{Appendix B}). The antiferromagnetic order with $\bm{Q}$ = ($\pi$, $\pi$) in the adjacent Mott insulating phase and a numerical study showing the enhancement of the $\bm{Q}$ = ($\pi$, $\pi$) fluctuations on approaching the BCMT~\cite{Watanabe2006,Kang2011,Tahara2008} suggest that the spin fluctuations probed by ($T_{1}T$)$^{-1}$ are highly weighted at $\bm{Q}$ = ($\pi$, $\pi$) near the Mott transition. These features suggest that the suppression of spin excitations on cooling occurs not on the entire Fermi surface but at particular $\bm{k}$-regions that affect AF spin fluctuations, which are very likely the crossing points of the Fermi surface and the zone boundary that are connected by $\bm{Q}$ = ($\pi$, $\pi$), as shown in Fig.~\ref{Fig1}(e). This notion is in accordance with the fact that the pseudogap-like behavior as observed here is absent in the pressure-driven metallic phase of the nearly triangular-lattice spin-liquid system $\kappa$-(ET)$_2$Cu$_2$(CN)$_3$~\cite{Shimizu2010}, which does not show AF ordering in its insulating state~\cite{Shimizu2003}.
\begin{figure}[!t]
\begin{center}
\includegraphics[width=8.6cm,clip]{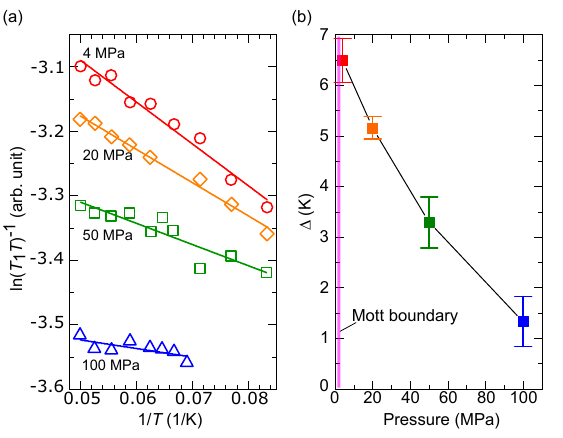}
\caption{(Color online) Pressure variation of phenomenological gap of ($T_{1}T$)$^{-1}$. (a) The Arrhenius plot of ($T_{1}T$)$^{-1}$ for various pressures. $(T_{1}T)^{-1}$ is approximated in the form of ($T_{1}T$)$^{-1}$ $\propto$ exp$(-\varDelta/T)$ between 12 K and 20 K. (b) The pressure dependence of $\varDelta$.
}
\label{Fig24} 
\end{center}
\end{figure}

To characterize the suppression of spin excitations on cooling, we provide rough estimates of the phenomenological gap, $\varDelta$, in the form of ($T_{1}T$)$^{-1}$ $\propto$ exp(-$\varDelta/T$) between 12 K and 20 K (see Fig.~\ref{Fig24}(a)). The $\varDelta$ value increases rapidly towards the BCMT (Fig.~\ref{Fig24}(b)), suggesting that the suppression of spin excitations on cooling is strongly connected with Mott localization.
We note that the deduced gap value should not be taken as it is because the temperature range of the fitting is comparable or higher than the gap value; nevertheless, the clear pressure dependence makes sense.


\section{Magnetic field study of $^{13}$C NMR}

In this section, we first show variation of the NMR spectra at ambient pressure on the verge of the Mott transition under magnetic fields ranging from 0.9 T to 18 T perpendicular to conducting planes. 
We successfully separated the the metallic phase from the insulating phase in spectra, which confirms the absence of symmetry breaking phases that competes with superconductivity, such as charge orders.
Next, we demonstrate the psuedogap-like anomalous suppression of spin excitations on cooling fades out in parallel with superconductivity with increasing magnetic field, indicating the ``pseudogap'' is a precursor of the superconductivity.
Our further investigation of different materials sees that the precursor is not amplitude fluctuations arising from low dimensionality but Cooper pair preformation promoted by electron correlations.

\subsection{NMR spectra under field variation}

\begin{figure}[!t]
\includegraphics[width=8.6cm,clip]{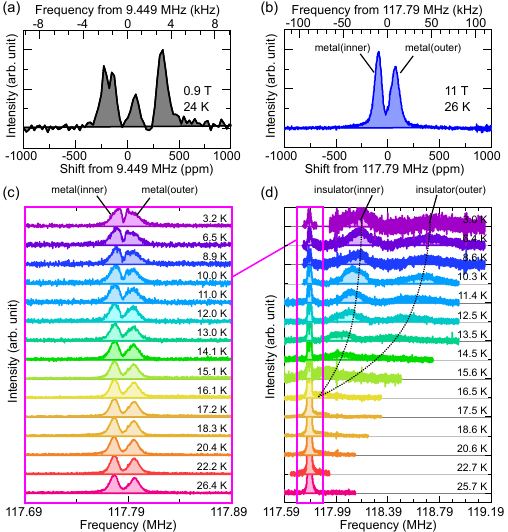}
\caption{
(Color online) Typical $^{13}$C NMR spectra of $\kappa$-dBr under perpendicular magnetic fields at ambient pressure. (a) 0.9 T (24 K) and (b) 11 T (26 K).
Temperature dependence of the NMR spectra of $\kappa$-dBr at ambient pressure under a magnetic field of 11 T applied perpendicular to the conducting layers: (c) a narrow frequency region, in which the metallic-phase spectra reside, and (d) a wide frequency range, which covers the insulating-phase spectra as well as the metallic-phase one.}
\label{Fig25} 
\end{figure}

In $^{13}$C NMR measurements with magnetic-field variation, we applied fields ranging from 0.9 T to 18 T parallel to the $b$ axis (out-of-plane) across the upper critical field of $H_{c2\bot}$(0 K) $\approx$ 10 T~\cite{Kwok1990}. 
To avoid experimental difficulty in tuning the resonant circuit (parts of which are inside the pressure cell) in frequencies over a twenty-fold range, the measurements were performed without a pressure cell at ambient pressure.
In $\mu_{0}H$ $\parallel$ $b$, four lines emerge owing to the two splitting mechanisms; (i) the inner-outer doublet and (ii) the Pake doublet, as explained above. The splitting of (i) is proportional to the magnetic field, whereas the splitting of (ii) is independent of the magnetic field; so, the spectral profile is largely varied by the magnetic field. In low magnetic fields, the splitting of (ii) is dominant or (i) can be comparable with (ii) so that the spectrum forms a quartet with two inner lines of small intensities~\cite{Mayaffre1995,DeSoto1996}. This is the case for the spectrum at 0.9 T as shown in Fig.~\ref{Fig25}(a) although two inner lines have merged into a central line in this particular magnetic field. In this case, each line has an indistinguishable relaxation rate owing to the admixture of the inner-site and outer-site relaxation rates, which are several-fold different. As a magnetic field is increased, the splitting of (i) becomes comparable to or larger than that of (ii); thus, the quartet comes to consist of an inner-site doublet and an outer-site doublet. At much higher magnetic fields (11, 15.5, and 18 T in the present case), each of the inner- and outer-site doublets loses the two-peak structure because the linewidth increases in proportion to the magnetic field (in other words, the splitting of the doublet decreases in ppm in inverse proportion to magnetic field), resulting in two lines that come from the inner and outer sites separately without mixture and thus have different relaxation rates ($T_{1})^{-1}$'s (see Fig.~\ref{Fig25}(b) for the spectrum at 11 T). From the spectral profiles, the difference between the inner- and outer-site shifts (splitting (i)) is found to be in a range of 200$-$300 ppm, and the splitting of the Pake doublet (splitting (ii)) is approximately 3 kHz.
\begin{figure*}[t]
\begin{center}
\includegraphics[width=17cm,clip]{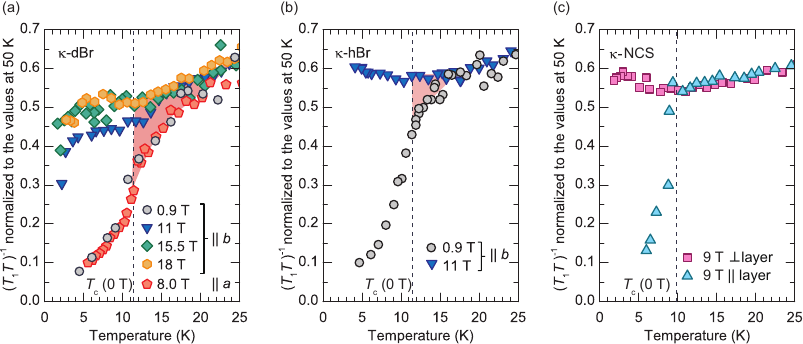}
\caption{(Color online) Material variation of ($T_{1}T$)$^{-1}$ under parallel and perpendicular magnetic-filed configurations. The temperature dependence of ($T_{1}T$)$^{-1}$ with different magnitudes and directions [$\parallel$ $b$ (out-of-plane) or $\parallel$ $a$ (in-plane)] of the magnetic field are shown for three $\kappa$-type superconductors with half-filled bands; (a) $\kappa$-dBr, (b) $\kappa$-hBr, and (c) $\kappa$-NCS, which are located further from the Mott transition in this order by chemical pressure. The values of ($T_{1}T$)$^{-1}$ are normalized to the values at 50 K (Appendix~\ref{Appendix D}.)}
\label{Fig26} 
\end{center}
\end{figure*}

At ambient pressure, the volume fraction of the metallic phase in the sample used is smaller (5$-$15 \% at 5 K) than under pressure; however, the spectral separation is clear.
Figures~\ref{Fig25}(c) and (d) show the temperature dependence of $^{13}$C NMR spectra of $\kappa$-dBr under a perpendicular field of 11 T at ambient pressure. The sharp double-peaked spectra coming from the metallic phase staying around 117.79 MHz and the broad double lines from the insulating phase that move toward higher frequencies on cooling are well separated without any continuous distribution. The domain sizes of the metallic and insulating phases are macroscopic (of the order of 100 $\rm \mu$m) according to the scanning micro-region infrared spectroscopy ~\cite{Sasaki2004}. Thus, the superconducting domains hold bulk properties in nature. Actually, the AC susceptibility measurements show a sharp superconducting transition at 11.6 K (see Appendix~\ref{Appendix C}).

As shown in Fig~\ref{Fig25}(c), the NMR spectra of the metallic phase in $\kappa$-dBr do not show any splitting, broadening, or complicated structures indicative of ``competing orders'', such as charge orders~\cite{Miyagawa2000}, charge density waves, or time-reversal symmetry breaking orders, on the verge of the Mott transition. We note that the absence of the competing order was confirmed at low temperatures well below $T_{\mathrm{c}}$(0 T), or under an extremely high magnetic field of 18 T, which corresponds to a magnetic field above tens of tesla in cuprate superconductors, considering the difference between the energy scales of the organics and the cuprates. 
Thus, thre present observation of the anomalous suppression of spin excitations is free from the intricate issue of competing or coexisting orders as observed in doped copper oxides.

\subsection{Magnetic field dependence of ($T_{1}T$)$^{-1}$}

Here, we show the magnetic field dependence of ($T_{1}T$)$^{-1}$. As mentioned above, just near ambient pressure, $\kappa$-dBr contains both metallic and insulating phases and the inner-site line of the insulating phase tends to near the outer-site line of the metallic phase at high temperatures. In addition, in high magnetic fields of 15.5 and 18 T, the whole spectra from the metallic phase were extended, respectively, over 150 and 200 kHz, which were not covered by the present NMR pulses. For these reasons, we used the inner-site lines for the evaluation of $T_{1}^{-1}$ in a high field and whole lines only for that in 0.9 T (see Appendix~\ref{Appendix D}).
As explained in the previous subsection, the spectral profile under the field $\parallel$ $b$ axis largely changes against field variation over the twenty-fold range. This causes a spurious field-dependence of the absolute value of ($T_{1}T$)$^{-1}$ due to the field-dependent mixing of inner-line $T_{1}^{-1}$and outer-line $T_{1}^{-1}$. 
Thus, ($T_{1}T$)$^{-1}$ is normalized to the values at 50 K, higher than the metal-insulator crossover temperature, where ($T_{1}T$)$^{-1}$ is expected not to depend on magnetic field (Appendix~\ref{Appendix D}). 

The normalized ($T_{1}T$)$^{-1}$ is clearly field-dependent at low temperatures (Fig.~\ref{Fig26}(a)). For a low perpendicular magnetic field, 0.9 T, a steep decrease in ($T_{1}T$)$^{-1}$ on cooling from well above $T_{\rm c}$ is evident. As the magnetic field is increased, however, that becomes less prominent at 11 T and largely suppressed at 15.5 and 18 T, which exceeds $H_{c2}$, although there remains a weak $T$-linearity. Thus, the recovery of the suppressed spin excitations above $T_{\rm c}$ proceeds in parallel with the destruction of the superconductivity. In addition, both of the superconductivity and the suppressed spin excitations are hardly affected by a parallel field of 8 T ($\parallel$ $a$ axis) as observed in Fig.~\ref{Fig26}(a). In the parallel magnetic field, the superconductivity is robust against the magnetic field because of the absence of the orbital depairing. The field-anisotropy common to the superconductivity and the anomalous spin excitations indicate their inseparable connection; namely, the latter is the manifestation of a superconducting precursor that persists up to twice as high as $T_{\rm c}$ (Fig.~\ref{Fig26}(a)).
The weak $T$-linear dependence under 18 T may suggest the presence of another pseudogap opened in incoherent carriers different from the preformed pairs~\cite{Kang2011, Gull2013, Imada2019}. As described below, the Knight shift also shows a decrease on cooling under 18 T.
Although evaluation and comparison of Knight shifts in different magnetic fields are not straightforward because of strongly field-dependent spectral shape as mentioned above, we show the normalized Knight shift for different magnetic fields in Fig ~\ref{Fig27} (see Appendix~\ref{Appendix E}). Above $T_{\rm c}$, the Knight shift under the parallel field of 8 T is not remarkably suppressed from that under the perpendicular field of 18 T, compared with the case of ($T_{1}T$)$^{-1}$ (Fig ~\ref{Fig27}). For example, at 14 K (11.5 K), the reduction in the Knight shift is only $\sim$ 1 \% ($\sim$ 14 \%), whereas the reduction in ($T_{1}T$)$^{-1}$ is $\sim$ 21 \% ($\sim$ 44 \%). 
Therefore, the present anomalous suppression of spin excitations should be ascribed to superconducting fluctuations with remarkable $\bm{q}$ dependence.
We note that the magnetic anisotropy of the present organic systems is negligibly small due to very small spin-orbit interactions, as is evident from nearly isotropic spin susceptibility and the $g$ factors of approximately 2.0. Thus, the field direction sensitivity of the anomalous decrease in ($T_{1}T$)$^{-1}$ is not explained by spin fluctuations separate from superconducting fluctuations. We note that the apparent temperature dependence of the Knight shift at 18 T, coinciding with the behavior of ($T_{1}T$)$^{-1}$, may suggest a strange metal with a pseudogap due to partially incoherent carriers~\cite{Kang2011, Gull2013, Imada2019}.

\begin{figure}[!ht]
\begin{center}
\includegraphics[width=6cm,clip]{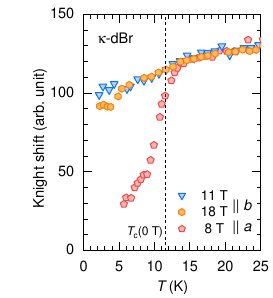}
\caption{(Color online) Temperature dependence of the normalized Knight shift for various magnetic fields. The procedure of the normalization is explained in Appendix~\ref{Appendix E}.
}
\label{Fig27} 
\end{center}
\end{figure}

\subsection{Material dependence of ($T_{1}T$)$^{-1}$}
To further look into the origin of the anomalous behavior of ($T_{1}T$)$^{-1}$ in $\kappa$-dBr, we investigated the material dependence of the phenomenon in question. Figure~\ref{Fig26}(b) shows ($T_{1}T$)$^{-1}$ for non-deuterated $\kappa$-(ET)$_2$Cu[N(CN)$_{2}$]Br ($\kappa$-hBr), which is situated further from the BCMT than $\kappa$-dBr is, at perpendicular magnetic fields of 0.9 T and 11 T. The results reproduce the essential features in previous reports~\cite{Kobayashi2014,Mayaffre1995}. A decrease in ($T_{1}T$)$^{-1}$ on cooling toward $T_{\rm c}$ is recognizable, however, it is less prominent than in $\kappa$-dBr. Figure~\ref{Fig26}(c) shows ($T_{1}T$)$^{-1}$ of $\kappa$-(ET)$_2$(NCS)$_2$ ($\kappa$-NCS, $T_{\rm c}$ = 10 K), located further away from the Mott boundary than $\kappa$-hBr, under parallel and perpendicular fields of 9 T. No anomalous behavior above $T_{\rm c}$ is evident, as also reported in Ref.~\cite{Kobayashi2014}. These results, in conjunction with the pressure dependence of ($T_{1}T$)$^{-1}$ of $\kappa$-dBr described above, indicate that the enhanced precursor to the superconductivity in $\kappa$-dBr originates from the electron correlations that are enhanced near the BCMT. The conventional low-dimensionality-driven fluctuations~\cite{Tinkham2004} comprised of the Aslamasov-Larkin, Maki-Thomson and density-of-states effects are ruled out by the absence of its signature in $\kappa$-NCS, which is the most highly two-dimensional among the three~\cite{Ito1991}; actually, the former two effects are shown not to affect ($T_{1}T$)$^{-1}$ in $d$-wave superconductivity~\cite{Kuboki1989,Larkin2005}. Thus, the superconducting precursor is most likely the preformation of phase-incoherent Cooper pairs.

\section{Discussion}
Here we explain step by step why the observed NMR behaviors indicate the preformed Cooper pairing above $T_{\mathrm{c}}$. Possible origins of the anomalous suppression of ($T_{1}T$)$^{-1}$ (and Knight shift) for metallic phase on cooling are following: (1) SC precursor, (2) magnetic fluctuations, (3) partial gap opening due to symmetry-braking order, and (4) $k$-dependent incoherence of quasiparticles due to proximity to the Mott transition. 
The first scenario (SC precursor effect) is consistent with the magnetic-field dependence of ($T_{1}T$)$^{-1}$; that is, the pseudogap-like behavior of ($T_{1}T$)$^{-1}$ above $T_{\mathrm{c}}$ and the superconducting condensate fade out in parallel under ascending magnetic fields with similar field-orientation dependence. 
As for the second scenario, the magnetic anisotropy of the present system is negligibly small due to very small spin-orbit coupling and this fact is inconsistent with remarkable magnetic field-orientation dependence of the pseudogap behavior of ($T_{1}T$)$^{-1}$. Thus, the scenario of magnetic fluctuations is unlikely.
The third scenario needs symmetry-breaking order with finite $\bm{q}$, such as charge order, density wave, or bond order~\cite{Tazai2021}. 
Such orders have to cause the inequivalence of the $^{13}$C nucleus in different sites and results in NMR spectral splitting and/or broadening. 
However, there is no such indication as seen in the result section.
As regards the fourth scenario, some theoretical studies suggest the formation of partial gap and Fermi arc near Mott transition~\cite{Kang2011, Gull2013, Imada2019}. 
However, this mechanism is not predominant origin of the anomalous behavior in ($T_{1}T$)$^{-1}$ because the pseudogap due to this scenario should be insensitive to the direction of a magnetic field. We nevertheless note that the field-independent residual linear temperature dependence of ($T_{1}T$)$^{-1}$ of $\kappa$-dBr at high magnetic fields (Fig~\ref{Fig26}(a)) might imply the opening a small partial gap causing $k$-dependent incoherence of quasiparticles.
Thus, the origin of the anomalous behavior of ($T_{1}T$)$^{-1}$ is ascribable to SC precursor.
In addition, the material dependence of ($T_{1}T$)$^{-1}$ clearly excludes the case of conventional amplitude fluctuations arising from low dimensionality and points to unconventional superconducting phase fluctuations meaning preformation of Cooper pairs enhanced near Mott transition.

\begin{figure}[h]
\begin{center}
\includegraphics[width=8.6cm,clip]{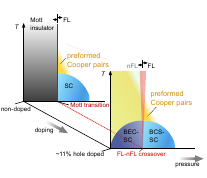}
\caption{(Color online) Schematic pressure-temperature phase diagrams of the doped~\cite{Suzuki2022} and the present non-doped Mott systems. FL and nFL denote Fermi liquid and non-Fermi liquid, respectively. Red solid and dotted lines indicate the lines of the first-order Mott transition and the sharp crossover from FL to nFL, respectively. The orange color depicts the region of the Cooper pair preformation.}
\label{Fig31} 
\end{center}
\end{figure}

The present observation that the AF correlations and the preformation of Cooper pairs are simultaneously enhanced near the BCMT (Figs~\ref{Fig23} and ~\ref{Fig24}) has a consistent explanation. As the system approaches the Mott metal-insulator transition, double occupancy on a site is strongly prohibited. Consequently, spins take on the localized nature, which leads to the enhancement of the ($\pi$, $\pi$) AF correlations. On the other hand, the increasing single-occupancy probability suppresses particle-density fluctuations and enhances the phase fluctuations due to the uncertainty principle so that the incoherent Cooper pairs are easily preformed~\cite{Emery1995}. Interestingly, recent Nernst effect study~\cite{Suzuki2022} for an organic doped Mott insulator $\kappa$-(ET)$_{4}$HgBr$_{2.89}$ indicates that bandwidth-controlled Bose-Einstein condensation(BEC)-to-Bardeen-Cooper-Schrieffer(BCS) crossover can occur around the exotic critical point (near the apex of the $T_{\rm c}$ dome) where the Fermi liquid sharply crosses over to the non-Fermi liquid and the double occupancy is prohibited~\cite{Oike2015}. Considering that this critical point can be regarded as a doped-system counterpart of the BCMT, it is reasonable to expect that $\kappa$-dBr is also on the verge of the BEC-BCS crossover, around which bosonic Cooper pairs are preformed at temperatures much higher than $T_{c}$. Moreover, comparing the pressure-temperature phase diagrams of the doped~\cite{Suzuki2022} and present non-doped systems, it is reasonable that the preformation of the Cooper pairs are immediately suppressed just after the Mott transition by pressure while $T_{\rm c}$ is not so, as illustrated in Fig.~\ref{Fig31}.
Interestingly, photo-induced effects related to superconducting fluctuations have been observed near the Mott transition~\cite{Kawakami2018,Kawakami2020,Buzzi2020,Buzzi2021}, and the photo-induced superconductivity is very sensitive to the distance of the system from the Mott transition~\cite{Buzzi2020,Buzzi2021}. Note that cluster dynamical mean field theory suggests that the onset temperature of local pair-formation monotonically increases on approaching the BCMT~\cite{Sordi2012a}.

Whether the superconducting gap symmetry of $\kappa$-(ET)$_{2}X$ is $d_{xy}$ (= $d_{X^{2}-Y^{2}}$ \footnote{Note that there are two natations describing gap symmetry of $\kappa$-(ET)$_{2}X$. The $d_{xy}$ ($d_{x^{2}-y^{2}} + s$) state in the folded Brillouiin zone (Fig.1(e)) is identical to the $d_{X^{2}-Y^{2}}$ ($d_{XY} +s$) state in the unfolded Brillouin zone}) or $d_{x^{2}-y^{2}} + s$ (= $d_{XY} + s$) has not been settled~\cite{Schmalian1998,Schrama1999,Izawa2001,Kuroki2002,Powell2007,Dion2009,Malone2010,Oka2015,Guterding2016,Guterding2016a,Cavanagh2019,Watanabe2017,Watanabe2019,Imajo2021} and argued as a subtle problem quite sensitive to the Fermi surface topology and/or the degree of dimerization of ET molecules ~\cite{Kuroki2002,Guterding2016a,Watanabe2017,Watanabe2019,Imajo2021}, where ($x$,$y$) and ($X$,$Y$) refer to the coordinates shown in Fig.~\ref{Fig1}(e). The present observation of the relation between the superconducting fluctuations and $\bm{Q}$ = ($Q_{X}$, $Q_{Y}$) = ($\pi$, $\pi$) AF fluctuations is consistent with the $d_{xy}$ paring scenario in the vicinity of the Mott transition. Similar to the copper oxide superconductors, the scattering with $\bm{Q}$ = ($\pi$, $\pi$) mediates the $d_{xy}$ paring~\cite{Schmalian1998,Powell2007}, which, in turn, opens a gap around the regions indicated by the gray circled area in Fig.~\ref{Fig1}(e), explaining why the anomalous behavior is particularly prominent in ($T_{1}T$)$^{-1}$ but not so in the Knight shift.

The present case is distinct from the widely recognized pseudogap in the copper oxides in that the pseudogap is clearly visible in the Knight shift~\cite{Timusk1999} and involves a sizable portion of the Fermi surfaces, as revealed by angle-resolved photoemission spectroscopy~\cite{Hashimoto2014}, and complex orders~\cite{Wu2018,Nie2014,Scheurer2018,Lee2014,Agterberg2019,Uchida2021,Ghiringhelli2012,Chang2012a,Wu2011,Xia2008,Lawler2010,Sato2017a,Agterberg2019}. Doped copper oxides are distinguished from the present system with a half-filled band in the following respects. First, copper oxides enter inside the strongly correlated region where double occupancy is strongly prohibited. Second, doping, which introduces vacancies, makes the charge degrees of freedom vital, whereas a half-filled band has no such room; in fact, the NMR spectra of $\kappa$-dBr show no signature of charge ordering at measured pressures, temperatures, and magnetic fields. Thus, the present observation of preformed pairs is addressed as an emergence under limited correlation and prohibited charge redistribution. When both restrictions are relaxed as a result of doping, the pseudogap behavior as observed in copper oxides might be induced. 
Doped organic Mott systems ~\cite{Oike2015,Oike2017,Suzuki2022,Imajo2021a} and organic Mott field-effect-transistor systems ~\cite{Kawasugi2016,Sato2017,Yamamoto2018,Kawasugi2019,Kawasugi2022} are useful to elucidate the issue in future.

\section{Summary}
In summary, the NMR experiments of quasi-two-dimensional organic superconductors under pressure control revealed that, above $T_{\rm c}$, the NMR relaxation rate is anomalously suppressed on cooling near the Mott transition. Experiments under the variation of mgnetic field in magnitude and orientation indicates that a superconducting precursor is responsible for the suppression of the relaxation rate on cooling.
By investigating three materials, which take different distance from the Mott transition, we could rule out the conventional amplitude fluctuations due to the low-dimensionality and confirmed the view of the relaxation suppression by preformed Cooper pairs with enhanced phase fluctuations. 
The present work that exploited physical and chemical pressures clarified that the Cooper pair preformation is saliently enhanced on the verge of Mott localization. NMR is a particularly advantageous microscopic probe for detecting incoherent precursory phenomena as revealed here.

\begin{acknowledgments}
We would like to thank K. Konishi, T. Yogi, D. Imamura, and F. Kagawa for experimental assistance on NMR measurement and M. Imada for fruitful discussion. This work was supported by JSPS KAKENHI (Grant Numbers: 20110002, 25220709, 24654101, 18H05225, 20K20890, 20K20894, 20KK0060 and 21K18144), the Mitsubishi Foundation (Grant Number: 202110014) and the U.S. National Science Foundation (Grant Number: PHYS-1066293). We thank for the hospitality of the Aspen Center for Physics and ICMR at the University of California in Santa Barbara.
\end{acknowledgments}

\appendix
\section{Materials and Methods}
\label{Appendix A}
\subsection{Sample preparation}
Single crystals of $\kappa$-(ET)$_2$Cu[N(CN)$_2$]Br ($\kappa$-hBr), its deuterated version ($\kappa$-dBr), and $\kappa$-(ET)$_2$Cu(NCS)$_2$ were grown using the conventional electrochemical method, in which two central carbon atoms in ET and deuterated (98 \%) ET molecules are enriched with $^{13}$C isotope by 99 \%, as shown in Fig.~\ref{Fig1}(c).
At the central carbons, the HOMO (highest occupied molecular orbital) has a high population; hence, through the large hyperfine coupling, the $^{13}$C nuclei probe the states of conduction electrons with high sensitivity, for example, compared with the $^{1}$H nuclei located on the edges of the ET. 
\subsection{NMR measurements}
The $^{13}$C NMR spectra were obtained by Fourier transformation of the quadrature-detected echo signals. Two types of pulse sequences were employed: the spin-echo pulse sequences of $(\pi/2)_{X}$-$(\pi)_{X}$ and the solid-echo pulse sequences of $(\pi/2)_{X}$-$(\pi/2)_{Y}$, where X and Y denote the axes in the rotational frame. For the pressure-dependence study, a coil wound around a sample was inserted in a pressure cell. Alternatively, for the magnetic-field-dependence study, the measurements were performed at ambient pressure without a pressure cell to avoid the experimental difficulty in tuning the resonant circuit, parts of which are inside the pressure cell, in frequencies over a twenty-fold range. 
\subsection{He-gas pressure}
To achieve fine control of the bandwidth of $\kappa$-dBr, we used the He-gas pressure technique. The NMR coil containing the sample was inserted into a pressure cell made of non-magnetic BeCu and compressed hydrostatically in a He pressure medium, which was directly compressed through a capillary tube by a gas-compressing system outside the cryostat. This technique allowed us to perform a continuous pressure-sweep while maintaining a nearly constant temperature, even at low temperatures, unless the He medium solidified.

\section{Evaluation of NMR relaxation-rate enhancement factor (the Korringa ratio)}
\label{Appendix B}
NMR relaxation rate measures the wave vector ($\bm{q}$)-summation of spin fluctuations weighted with the squared form factor $|A(\bm{q})|^{2}$ over the first Brillouin zone, as expressed by ($T_{1}T$)$^{-1}$ $ \propto$ $\sum$$_{\bm{q}}$$|A(\bm{q})|^{2}$$\chi$''($\bm{q}$,$\omega_{\rm{NMR}}$)/$\omega_{\rm{NMR}}$, where $\chi$''($\bm{q}$,$\omega$) is the imaginary part of the dynamic spin susceptibility and $\omega_{\rm{NMR}}$ is the resonance frequency of NMR measurement. In the present case, the form factor $A$($\bm{q}$) of the $^{13}$C site located at the midst of the molecular orbital in the ET molecule is $\bm{q}$-independent and thus ($T_{1}T$)$^{-1}$ $\propto$ $\sum$$_{\bm{q}}$ $\chi$''($\bm{q}$,$\omega_{\rm{NMR}}$). In such a case, the Korringa ratio $\mathcal{K}_{\alpha}$ evaluated from the experimental values of ($T_{1}T$)$^{-1}$ and Knight shift $K$ provides information on the $\bm{q}$ profile of $\chi$''($\bm{q}$,$\omega$). In case of the isotropic hyperfine coupling, it is well known that $\mathcal{K}_{\alpha}$ is given by 
$\mathcal{K}_{\alpha}$ = $(\hbar$/4$\pi$$k_{\rm{B}}$)($\gamma_{\rm{e}}$/$\gamma_{\rm{n}}$)$^{2}$($T_{1}T$)$^{-1}$$K^{-2}$, where $\hbar$ is Plank's constant, $k_{B}$ Boltzmann's constant, and $\gamma_{\rm{n}}$ and $\gamma_{\rm{e}}$ are the gyromagnetic ratios of nuclear spin and electron spin, respectively. 
In the present case, however, the hyperfine coupling is anisotropic and expressed by a tensor; the principal values of the hyperfine coupling tensor of the inner $^{13}$C are $a_{\tilde{x}\tilde{x}}$ = -1.4 kOe/$\mu_{\rm{B}}$, $a_{\tilde{y}\tilde{y}}$ = -3.3 kOe/$\mu_{\rm{B}}$ and $a_{\tilde{z}\tilde{z}}$ =10 kOe/$\mu_{\rm{B}}$ with $\tilde{x}$, $\tilde{y}$ and $\tilde{z}$ axes indicated in Fig.~\ref{Fig1}(b) according to Refs.~\cite{Miyagawa2004a,DeSoto1995}, where $\mu_{\rm{B}}$ is the Bohr magneton. In this case, the Korringa ratio is expressed by
$\mathcal{K}_{\alpha}$ = $\beta(\zeta,\eta)$($\hbar$/4$\pi$$k_{\rm{B}}$)($\gamma_{\rm{e}}$/$\gamma_{\rm{n}}$)$^{2}$($T_{1}T$)$^{-1}$$K^{-2}$, where $\beta(\zeta,\eta)$ is given by 
\begin{multline}
\beta(\zeta,\eta) = 2[a_{\tilde{x}\tilde{x}}\sin^{2}\zeta\cos^{2} \eta \\
+ a_{\tilde{y}\tilde{y}}\sin^{2}\zeta\sin^{2}\eta + a_{\tilde{z}\tilde{z}}\cos^{2}\zeta]^{2} \\
/[a_{\tilde{x}\tilde{x}}^{2}(\sin^{2}\eta+\cos^{2}\zeta\cos^{2}\eta) + \\
a_{\tilde{y}\tilde{y}}^{2}(\cos^{2}\eta+\cos^{2}\zeta\sin^{2}\eta) + a_{\tilde{z}\tilde{z}}^{2}\sin^{2}\zeta]. \notag
\end{multline}
$\zeta$ is the angle between the external field and the $\tilde{z}$-principal axis and $\eta$ is the polar angle measured from the $\tilde{x}$-principal axis in the $\tilde{x}\tilde{y}$ plane of Fig.~\ref{Fig1}(c).
The substitution of the experimental data in Figs.~\ref{Fig23}(a) and (b), and the experimental angles, $\zeta$ = 55$^{\circ}$ and $\eta$ = 43$^{\circ}$ to the form of $\mathcal{K}_{\alpha}$ yields $\mathcal{K}_{\alpha}$ $\sim$ 8-12 in the metallic phase (e.g. $\mathcal{K}_{\alpha}$ = 8.1 for $P$ = 100 MPa and $T$ = 15 K), which greatly exceeds unity.
It means that ($T_{1}T$)$^{-1}$ $\propto$ $\chi$''($\bm{q}$,$\omega_{\rm{NMR}}$) is overwhelmingly contributed by components with $\bm{q}$ $\neq$ 0, namely antiferromagnetic fluctuations.

\section{Characterization of superconductivity in $\kappa$-dBr at ambient pressure}
\label{Appendix C}
\begin{figure}[!htb]
\begin{center}
\includegraphics[width=6cm,clip]{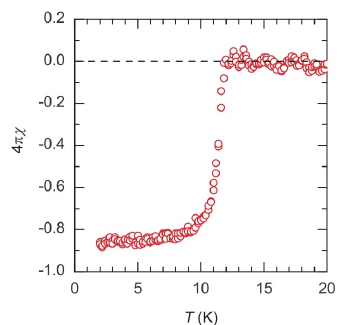}
\caption{(Color online) 
Temperature dependence of the AC susceptibility of $\kappa$-dBr at ambient pressure. Demagnetizing field was corrected (see Appendix~\ref{Appendix C}).
}
\label{Fig28} 
\end{center}
\end{figure}

In the present study, the field dependence of $^{13}$C NMR measurements was performed at ambient pressure, where superconducting and Mott-insulating phases are coexistent due to the first-order Mott transition. From the NMR spectral profile, the volume fraction of the superconducting phase is approximately 15 \%. To examine the homogeneity of $T_{\mathrm{c}}$ in the superconducting domains, we measured AC susceptibility of the $\kappa$-dBr crystal used in the NMR measurements. Figure ~\ref{Fig28} shows the AC susceptibility measured with the AC filed of 17 Hz in frequency and 0.005 gauss in amplitude applied perpendicular to the conducting plane. The demagnetizing effect is corrected with the demagnetization factor of 0.58, which was determined by the measurements of Sn shaped into the same geometry as the $\kappa$-dBr crystal. A sharp transition at 11 K is evident and there is no feature indicative of distribution of $T_{c}$. The diamagnetic susceptibility is nominally 85 \% in shielding diamagnetism, which largely overestimates the superconducting volume fraction as expected. 

\section{Spurious field dependence of $T_{1}^{-1}$}
\label{Appendix D}
\begin{figure}[t]
\begin{center}
\includegraphics[width=8.6cm,clip]{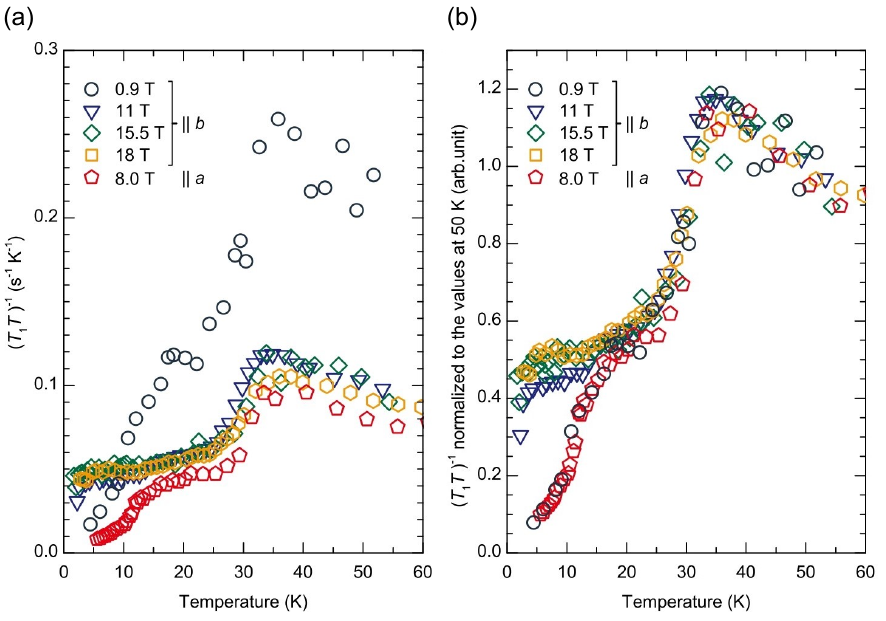}
\caption{(Color online) 
Temperature dependence of ($T_{1}T$)$^{-1}$ for various magnetic fields. (a) Raw values of ($T_{1}T$)$^{-1}$. (b) ($T_{1}T$)$^{-1}$ normalized to the values at 50 K.
}
\label{Fig29} 
\end{center}
\end{figure}

As explained in detail in the main text, the NMR spectrum under the field perpendicular to the layers ($\parallel$ $b$ axis) is comprised of a quartet that comes from the nuclear-dipole coupling of two adjacent $^{13}$C sites (`inner' and `outer' sites) with different Knight shifts in the ET molecule. At high fields (11, 15.5, 18 T), the dipolar coupling is secondary compared with the difference of the Knight shift so that the `inner' doublet and `outer' doublet are well separated; thus, the relaxation rate of the inner site is investigated as in the parallel-field case. At low fields (0.9 T), however, the two doublets are mixed up and the relaxation rates for the inner and outer sites get inseparable; therefore, the relaxation rate measured at low fields yields an average of the rates for the two sites. This situation gives a spurious field-dependence of the ($T_{1}T$)$^{-1}$ values.
Figure ~\ref{Fig29}(a) shows the temperature dependence of ($T_{1}T$)$^{-1}$ for various magnetic fields. Except at low temperatures (below 20 K), where the system exhibits superconductivity and anomalous spin excitations, the ($T_{1}T$)$^{-1}$ values at 11, 15.5 and 18 T are nearly coincident with each other in magnitude, because they correspond to the values for the inner-site. At 0.9 T, the magnitude of ($T_{1}T$)$^{-1}$ becomes greater because of the averaging of the inner- and outer-site values, as described above. 
To remove this spurious magnetic field dependence, the values of ($T_{1}T$)$^{-1}$ are normalized to the values at 50 K, where $\kappa$-dBr is in a paramagnetic Mott insulating phase and the magnetic field dependence of ($T_{1}T$)$^{-1}$ is expected to be negligible. As shown in Fig. ~\ref{Fig29}(b), the temperature dependences of the normalized values for different magnetic fields nearly coincide above 20 K. The steep decreases at approximately 30 K reflect a sharp crossover from the high-temperature Mott insulating phase to the low-temperature metallic phase, confirming that the metallic phase at low temperatures is properly captured by the present measurements and analyses. 
We also note that by using the so-called solid-echo pulse sequence, which is effective (ineffective) for refocusing the dipole-split (spin-shifted) lines, the quartet coming from the metallic phase with small spin shift is selectively picked up at low magnetic fields.

\section{Magnetic field dependence of the Knight shift for $\kappa$-dBr}
\label{Appendix E}
\begin{figure}[h]
\begin{center}
\includegraphics[width=6cm,clip]{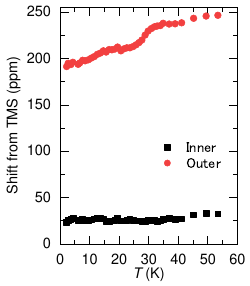}
\caption{(Color online) 
Temperature dependence of the inner-site and outer-site spectral shifts from TMS under the perpendicular field of 11 T.
}
\label{Fig30} 
\end{center}
\end{figure}
It is not straightforward to reveal the field dependence of the Knight shift because of the following reasons:
(1) an anisotropic hyperfine coupling tensor makes it difficult to compare the absolute values of the Knight shift measured under parallel and perpendicular fields,
(2) the spectral profile under perpendicular fields is very field-sensitive and somewhat complicated, because the relative strength of nuclear dipole coupling and the shift difference between the adjacent $^{13}$C nuclei is varied under field variation, making the determination of shift not straightforward,
(3) for a low perpendicular field of 0.9 T, poor signal-to-noise ratio as well as (1) and (2) makes it further ambiguous to determine the Knight shift and,
(4) the origin of the Knight shift under perpendicular magnetic fields is not known because the nearly complete singlet ground state, which is realized in parallel fields, is not attained in this field configuration.
Notwithstanding these difficulties, however, we have tried to reveal the field dependence of the Knight shift under the following restrictions.
First of all, we could not discuss the Knight shift under the low perpendicular field of 0.9 T by the reasons, (1), (2) and (3). At much higher fields ($\geq$ 11 T), the nuclear dipole coupling is not influential for extracting the Knight shift and thus we focus on the Knight shift under a parallel field of 8T and high perpendicular fields of 11 and 18 T. Note that the behavior of ($T_{1}T$)$^{-1}$ at the parallel field of 8 T is nearly the same as that at the perpendicular filed of 0.9 T, indicating the negligible field effect in both cases; thus, the Knight shift is expected to behave similarly in the two cases.
To compare the Knight shift values under perpendicular and parallel fields, we need to take into consideration the anisotropy of the hyperfine coupling tensor (the issue of (1)) and make a reasonable assumption on the shift origin under perpendicular fields (the issue of (4)). 
For this purpose, we converted the measured shift under the perpendicular field, $S_{\perp}$, to the normalized Knight shift, $K_{\perp}$ = $\alpha$ $\times$ $S_{\perp}$ + $\beta$ which is to be compared with the shift under parallel fields.
The parameter $\beta$ is equal to the ratio of the hyperfine coupling constants for the two field geometries,
$A_{\parallel a, outer}/A_{\parallel b, outer}$, which is calculated to be 1.47 $\pm$ 0.10 using the data in Ref.~\cite{DeSoto1995}.
The parameter $\beta$, which corresponds to the chemical shift (or the origin of the Knight shift), is determined such that the Knight shift averaged over 20-23 K under a perpendicular field is equal to the corresponding value of the Knight shift under the parallel field of 8 T.
Note that the hyperfine coupling constant of the inner site for a perpendicular field is much smaller than that of the outer site. Indeed, the measured shift of the inner-site spectrum for a perpendicular field is too small to be discussed (Fig.~\ref{Fig30}); hence, we focused on the outer-site spectral shift. 
Also note that, at 18 T, the data on the measured spectral shift were scattered owing to the instability of the applied field. To remove this scattering, we investigated the difference between the inner-site and outer-site shifts at each temperature. As mentioned above, for a perpendicular filed, the hyperfine coupling constant of the inner sites is much less than that of the outer sites; thus, the difference between the two values is approximately equal to the outer-site Knight shift and, even if the inner shift has a temperature-dependent finite value, the inner-outer shift difference, which should proportional to the shift itself, is normalized at 20-23 K, making no influence on the present discussion.
As a result, we found that the Knight shift under a parallel field of 8 T is much less suppressed from that under high perpendicular fields than ($T_{1}T$)$^{-1}$ is, as shown in Fig.~\ref{Fig27}.

\bibliography{Lib}
\end{document}